\documentclass[smallcondensed]{svjour3}     

\smartqed  
\usepackage{float}
\usepackage{graphicx}
\usepackage{bm}
\usepackage{xcolor}

\usepackage{amsmath}
\usepackage{amsfonts}
\usepackage{amssymb}
\usepackage{amstext}
\usepackage{amsbsy}
\usepackage{amsopn}
\usepackage{amscd}
\usepackage{amsxtra}

\usepackage[french,english]{babel}
\usepackage[colorlinks=true,allcolors=blue]{hyperref}
\usepackage{booktabs}

\newcommand{\Hilbert}{\mathcal{H}}
\newcommand{\Mc}[1]{\mathcal{#1}}

\newcommand{\setN}{\mathbb{N}}
\newcommand{\setR}{\mathbb{R}}

\newcommand{\setC}{\mathbb{C}}
\newcommand{\Id}{\mathbb{I}}

\newcommand{\bra}[1]{\langle #1 |}
\newcommand{\ket}[1]{| #1 \rangle }
\newcommand{\braket}[2]{\langle #1| #2 \rangle }
\newcommand{\ii}{\textsl{i}}

\renewcommand{\i}{\mathrm{i}}

\newtheorem{hyp}{Hypothesis}

\begin{document}

\title{A Model of Spinfoam Coupled with an Environment}

\author{Quentin Ansel} 
\institute{Institut UTINAM (CNRS UMR 6213, Université de Bourgogne-Franche-Comté, Observatoire de Besançon), 41~bis Avenue de l’Observatoire, BP1615, 25010 Besançon cedex, France.
              \email{quentin.ansel@univ-fcompte.fr}  
}

\date{Received: date / Accepted: date}

\maketitle

\begin{abstract}
In this paper, an open quantum system theory for spinfoams is developed. This new formalism aims at deriving an effective Lindblad equation to compute the reduced dynamics of a quantum gravitational field. The system parameters are determined from numerical ab initio calculations, based on the spinfoam formalism. This theoretical proposal is illustrated by means of examples. The decoherence effect can induce the relaxation of the quantum gravitational state toward a sate of a small area. This is analogue to the well-known Purcell relaxation of QED, for which the qubits are replaced by the spin-network representation of the gravitational field. Some thermodynamic properties of these systems are computed, and several issues with the thermal time hypothesis are underlined. Moreover, the results suggest that further approximations can be performed to study reduced dynamics of quantum space-time.

\end{abstract}

\section{Introduction}

Quantum gravity is one of the main issues in modern physic, and several concurrent theories try to provide answers to this difficult physical problem. Among the different approaches, Loop Quantum Gravity (LQG)  \cite{gambini_loops_1996,rovelli_quantum_2007,ashtekar_general_2014,rovelli_space_2018} and its covariant formulation \cite{rovelli_covariant_2014,rovelli_simple_2011,perez_spin-foam_2013} have received an increasing interest during the last 25 years, due to many theoretical successes (well-defined quantization, derivation of physically meaningful quantities, 4D Lorentizan path integrals, semi-classical limit, numerous predictions in black hole physics and cosmology, experimental predictions,... see \cite{rovelli_covariant_2014,barrau_loop_2014,cai_testing_2014,perez_black_2017,christodoulou_transition_2017,dambrosio_end_2020} for non-exhaustive applications). The original idea of LQG consists of expressing the gravitational field with functions of small loops that can be thought as space-time paths, instead of a metric modified by gravitational waves. Up to this date, computation of LQG dynamics have been the most successful with the spinfoams formalism, a specific approach based on path integral concepts. Thanks to new numerical methods \cite{dona_numerical_2018,dona_numerical_2019}, it is now possible to evaluate numerically the path integral. Unfortunately, it is a very demanding task, and only foams of small dimensions can be considered.

The aim of this paper is to introduce an open quantum system approach \cite{breuer_theory_2007} for the study of\textit{ a portion of space-time} interacting with its surrounding environment. The final goal being to simplify the study of quantum space-time dynamics. This approach is particularly relevant for models of space-time monitored by an exterior observer (an ensemble of fields, which act as sources of decoherence).
The application of open quantum system theory to quantum gravity has been considered in reference~\cite{feller_entanglement_2017}. In this work we take a different path whereby an effective system is defined from the adiabatic elimination method of the  Lindblad equation \cite{azouit_adiabatic_2016}. The resulting effective Lindblad equation is composed of several damping operators for which the coupling constant depends on a spinfoam probability transition. These terms can be computed
with ab initio numerical simulation of the path integral \cite{dona_numerical_2018}. The corresponding effective system is analogous to an ensemble of qubits in interaction with a leaky cavity. Then, the system can be studied using state-of-the-art tools of open quantum system theory \cite{breuer_theory_2007}. 
This theoretical proposal allows us to tackle several problems from a novel point of view, such as the transition of geometries \cite{christodoulou_transition_2017}, or the meaning of time in quantum gravity \cite{paetz_analysis_2010}.

This paper is organized as follows: In section \ref{sec1}, spinfoam theory is introduced, and the concept of sub-spin-network is developed. In section \ref{sec:effective_model}, the proposal of open spinfoam theory is presented and it is explained how parameters are determined, from path integral calculations. In section \ref{sec:decoherence of the gravitational field}, we study dynamics of some examples, and a few of their properties. This section finishes with a short discussion of the thermal time hypothesis applied to the effective model. Finally, a conclusion is given in section \ref{sec:conclusion}.

\section{Spin-networks and Spinfoams}
\label{sec1}

This section is devoted to a brief introduction to spinfoam theory and quantum gravity in the LQG framework.  First, some general concepts are given, and the idea of a sub-spin-network is introduced and formalized in a second step. Further mathematical definitions are given in \ref{sec:math_def}. 

\subsection{General definitions}
A quantum gravitational state is a function of variables that describes the classical geometry of a space-like sub-manifold of the space-time manifold \cite{baez_introduction_1999,rovelli_quantum_2007,rovelli_covariant_2014}. In LQG, it is convenient to express such a state with a spin-network $(\Gamma, \chi,\imath)$ (see \ref{sec:math_def} for further details). Assuming a discretization of the sub-manifold with a finite dimensional lattice $\Gamma$, composed of $N$ nodes and $L$ links, the Hilbert space of gravitational states is:
\begin{equation}
\Mc H_\Gamma = L_2 (\mathfrak H ^ L / \mathfrak{H}^N).
\end{equation}
The scalar product is defined with the Haar measure on $\mathfrak H$. In the specific examples considered below $\mathfrak H = SU(2)$.
From this construction, it is possible to define operators describing physical observable. A standard example in LQG is the area operator associated with a link \cite{rovelli_quantum_2007,rovelli_covariant_2014}:
\begin{equation}
\begin{split}
\hat A_l \ket{j_l} & = A_l \ket{j_l} \\
&=\frac{8 \pi \gamma_I \hbar G}{c^3} \sqrt{j_l(j_l+1)} \ket{j_l},
\end{split}
\label{eq:face_area}
\end{equation}
where $\ket{j_l}=\chi^{j_l}_l$ is the $SU(2)$ character in the $j_l$ representation, associated with the link $l$ of the graph $\Gamma$. Additionally $\gamma_I$ is the Immirzi parameter, $\hbar$ is the reduced Planck constant, $G$ is Newton's gravity constant, and $c$ is the light speed. In the following, we use plank units $G=\hbar=c = 1$.

Dynamics of the theory are given by a function that maps a spin-network to another one. This mapping is made explicit through a path-integral:
\begin{equation}
W(\psi_{out},\psi_{in}) = \int_{\psi_{in}}^{\psi_{out}} D[\phi] e^{\ii S[\phi]}.
\label{eq:path_integral}
\end{equation}
An expression of equation~\eqref{eq:path_integral} can be derived from the spinfoam formalism (see \ref{sec:math_def} and References  \cite{baez_introduction_1999,rovelli_quantum_2007,rovelli_covariant_2014,livine_short_2011,rovelli_simple_2011} for further definitions):
\begin{equation}
W(\psi_{out},\psi_{in}) = \sum_{F:\psi_{in}\rightarrow \psi_{out}} \frac{\lambda ^{V}}{\text{sym}(F)} Z(F),
\end{equation}
with $F$ a spinfoam $(\tilde{\Gamma}, \tilde{\chi},\tilde \imath)$ defined on a 2-complex $\tilde \Gamma$ made of $V$ vertices. $\lambda$ is a coupling constant. The sum is over all foam bounded by the spin-network $\psi_{in}$ and $\psi_{out}$.

All examples given in the main text of this paper are based on the EPRL model~\cite{engle_lqg_2008,rovelli_simple_2011,livine_short_2011,rovelli_covariant_2014}. In this formalism, $Z(F)$ is given by $\braket{Z}{\psi^*_{in} \otimes \psi_{out}} = \int_{SU(2)^L} Z(h_l)^*. ~ \psi^*_{in} \otimes \psi_{out} (h_l) ~ dh_l $, where $\psi^*_{in} \otimes \psi_{out}$ is a spin-network at the boundary  of the spinfoam. This special spin-network is defined by the disjoined union of the underlying graphs $\Gamma_{in}$ and $\Gamma_{out}$ . The weight function $Z(h_l)$ is defined by:
\begin{equation}
\begin{split}
Z(h_l) =& \int_{SL(2,\setC)^{2(E-L)-V}}dg_{ev} \int_{SU(2)^{\Mc V-L}}  dh_{ef}\\
& \sum_{j_f} \prod_f (2j_f+1) \chi^{\gamma_I j_f,j_f}\left( \prod_{e\in \partial f} g_{ef}^{\epsilon_{lf}}\right)\prod_{e \in \partial f} \chi ^{j_f} (h_{ef}),
\end{split}
\end{equation}
with $h_l, h_{ef} \in SU(2)$, $g_{ev} \in SL(2,\setC)$, and $g_{ef} \equiv g_{es} h_{ef} g_{et}^{-1}$ for internal edges, and $g_{ef}\equiv h_l$ for boundary edges. $ \chi^{\gamma_I j_f,j_f}$ is the $SL(2\setC)$ character in the unitary representation $(\gamma_I j_f,j_f)$, and $\chi^{j_f}$ is the $SU(2)$ character in the $j$ representation. $\gamma_I \in \setR$ is the Barbero-Immirzi parameter. Furthermore, $V$ is the number of vertices, $E$ is the number of edges, $L$ is the number of links, and $N$ is the number of nodes.
 
In the following, we \textit{assume} that $W(\psi_{out},\psi_{in})$ is well defined, finite, and that a spin-network Hilbert space $\Hilbert_\Gamma$ can be spanned by a countable set of orthogonal spin-networks. We introduce $\{\psi_i \equiv \ket{i} \}$, a basis of $\Hilbert_\Gamma$ such that $\braket{i}{j}=\delta_{ij}$. In particular, all computations are based on non-divergent spinfoams.

\subsection{Sub-spin-networks}
In this paper, we are not interested in a process governing the entire space, but on the dynamics of restricted areas. For that purpose, we introduce sub-spin-networks. Their goal is to introduce a notion of partial trace over space degrees of freedom.

\begin{definition}
A sub-spin-network $\psi'=(\Gamma',\chi,\imath)$ of a spin-network $\psi=(\Gamma,\chi,\imath)$ is given by a portion of $\psi$ defined by the connected complex $\Gamma' \subset \Gamma$. The complementary of a sub-spin-network $\bar \psi'$ is defined similarly, such that $\Gamma = \Gamma' \cup \bar \Gamma'$. 
\end{definition}
\begin{figure}\sidecaption
\includegraphics[width=8cm]{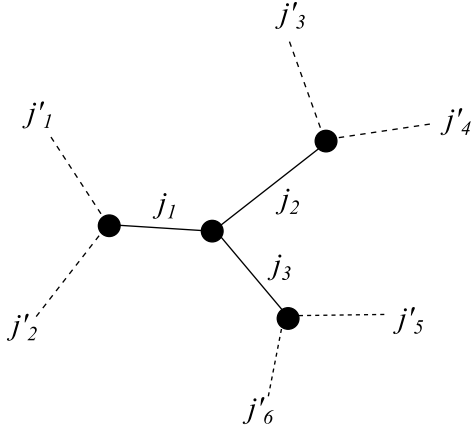}
\caption{An example of sub-spin-network. For simplicity, the example if taken from the Euclidean 3D theory. Solid lines represent links of $\Gamma '$, while dashed lines represent surrounding links of $\bar \Gamma '$. Only a portion of $\bar \Gamma'$ is drawn.}
\label{fig1}
\end{figure}
An example of sub-spin-network is sketched in figure~\ref{fig1}. Given a basis of $\Mc H_\Gamma$, a set of orthogonal sub-spin-networks compatible with $\Gamma '$ can be deduced, and a Hilbert space $\Mc H_{\Gamma '}$ can be defined.
\begin{definition}
The union map of a sub-spin-network with its complementary is defined such that: $\psi' \sqcup \bar\psi' : (\psi' ,\bar\psi') \rightarrow \psi$. Since the union of $\psi'$ and $\bar \psi'$ make sense only with the prior knowledge of $\psi$, the definition is extended to any state of $\Mc H_{\Gamma'}$ and $\Mc H_{\bar \Gamma'}$ with $\psi = P_{\Mc H_{\Gamma'}\otimes \Mc H_{\bar \Gamma'} \rightarrow \Mc H_{\Gamma}}(\psi' \otimes \bar \psi')$.
\end{definition}

Precisely, with the decomposition of the two sub-spin-networks on basis states:
\begin{equation}
\left( \sum_{n} c_n \ket{n}\right) \sqcup \left( \sum_{k} c_k' \ket{k}\right) = \sum_{n,k} c_n c'_k \ket{n}\sqcup\ket{k} = \sum_{i=(n,k)} \tilde c_i \ket{i},
\end{equation}
with $\tilde c_i = c_n c'_k$ and $ \ket{i}=\ket{n}\sqcup\ket{k} $ if $(n,k)$ is compatible with the spin-network structure of the full graph. Otherwise $\ket{n}\sqcup\ket{k} = 0 \ket{n,k}$. 

The union map must be used with caution since two arbitrary sub-spin networks cannot be connected in general. This can be observed with the figure~\ref{fig1}. Unconnected links must match the nodes where they are supposed to be inserted. This impacts possible values of $j$ (e.g. $j_1$ is constrained by $j_1'$ and $j_2'$). Such constraints are implicit in the union map. If two sub-spin-networks cannot be glued together, they contribute with a weight 0. In applications, it is necessary to specify a class of sub-spin-networks that must be adapted to the system under study. This step is similar to the work done in other studies \cite{christodoulou_transition_2017,christodoulou_planck_2016,dambrosio_end_2020}, for which a precise spin-foam structure, and a class of boundary state is chosen for the computations.

\section{Spin-network interacting with an environment}
\label{sec:effective_model}

Currently, the most rigorous theory of open quantum system is based on the Lindblad equation or similarly on the stochastic Schrodinger equation. The goal of this section is to introduce a theory of open quantum systems for a sub-spin-network. The theory presented below uses an approach inspired by the adiabatic elimination of a master equation \cite{azouit_adiabatic_2016}, which has two distinct time scales. First, the adiabatic elimination approach is introduced, and then, the idea is extended to spinfoams.

\subsection{Adiabatic elimination for open quantum systems}
\label{sec:Adiabatic elimination for open quantum systems}

In the following paragraphs, we consider a quantum system equipped with a finite dimensional Hilbert space $\Mc H$  (a generalization to infinite dimensional space is possible). We denote by $\Mc D$ the compact convex set of density operators $\rho$ on $\Mc H$. 
A general master equation is given by~\cite{lindblad_generators_1976,breuer_theory_2007}:
\begin{equation}
\begin{split}
\frac{d\rho(t)}{dt} & = -\ii [H(t),\rho(t)] + \sum_{n} \underbrace{R_n(t) \rho(t) R_n^\dagger(t) -\frac{1}{2}   \left( R_n^\dagger(t)R_n(t) \rho(t) + \rho(t) R_n^\dagger(t)R_n(t) \right)}_{D_{R_n(t)} [\rho(t)]} \\
& =  -\ii [H(t),\rho(t)] + \sum_n D_{R_n(t)} [\rho(t)] \\
& = \Mc L (t) \rho(t) ,
\end{split}
\end{equation}
with $\Mc L$ the Lindblad super operator. Coherent dynamics are given by the Hamiltonian $H$, and incoherent dynamics are given by $D_{R_n}$. These additional terms are sources of decoherence, and they generally come from the interaction with an environment which has many degrees of freedom. Notice that the number of damping operators depends on the system and its environment. Markovian dynamics are obtained when all $D_{R_n}$ are constant over time. Otherwise, we may have non-Markovian dynamics.

A Lindblad equation with two-time scales is given by:
\begin{equation}
\frac{d\rho}{dt} = \left(\Mc L_0 + \epsilon \Mc L_1 \right) \rho,
\label{eq:two-time-scale-lindblad-eq}
\end{equation}
with $\epsilon$ a small positive real number. We assume a convergence toward a stationary regime of the system defined by \eqref{eq:two-time-scale-lindblad-eq}, when $\epsilon = 0$. The set $\Mc D_0$ of steady state is given by:
\begin{equation}
\Mc D_0 = \{ \rho \in \Mc D ~|~ \Mc L_0 \rho = 0 \}.
\end{equation}
Equivalently, this set is described by:
\begin{equation}
\Mc D_0 = \{ \lim_{t \rightarrow \infty} e^{t \Mc L_0} \rho ~|~  \rho \in \Mc D \}.
\end{equation}
We also assume that $\Mc D_0$ coincides with a set of density operators with support in a subspace $\Mc H_0$ of $\Mc H$. Since $\epsilon$ is a small parameter, compared with other characteristic frequencies of the system, it is expected that dynamics follow mainly the unperturbed ones, and the perturbation is restricted to modify the system state around $\Mc D_0$. Then, it is interesting to describe the dynamics with an effective master equation describing the evolution of the subsystem \cite{azouit_adiabatic_2016}.

\begin{proposition}
\label{prop:adiabatic_elimination}
Let $\rho_e \in \Mc D_0$, the effective density matrix of the system governed by the Lindblad equation \eqref{eq:two-time-scale-lindblad-eq}. We assume that $\Mc L_1$ is such that:
\[
\Mc L_1 \rho \equiv  \sum_{n = 1}^{dim \Mc H_0} \left( P_n \rho P_n ^\dagger -\frac{1}{2} P_n^\dagger P_n \rho - \frac{1}{2} \rho P_n^\dagger P_n \right),
\]
with $P_n = \ket{n}\bra{n}$ a projector on an orthogonal basis of $\Mc H_0$. Additionally, we assume that unperturbed dynamics are given by: $U_0\rho(0)=\lim_{t \rightarrow \infty}e^{t\Mc L_0}\rho(0) \equiv \sum_\mu M_\mu \rho(0) M_\mu ^\dagger$. The first order expansion in $\epsilon$ of this equation around $\Mc D_0$ is a Lindblad equation of the form:
\[
\frac{d \rho_e}{dt} = \epsilon \sum_{n,m = 1}^{dim \Mc H_0} \kappa_{nm}\left( R_{nm} \rho_e R_{nm}^\dagger -\frac{1}{2} R_{nm}^\dagger R_{nm} \rho_e - \frac{1}{2} \rho R_{nm}^\dagger R_{nm} \right)
\]
with $R_{nm} =  \ket{n}\bra{m}$, and $\kappa_{nm} = \sum _\mu |M_{\mu,nm}|^2$. 
\end{proposition}
The proof is given in \ref{sec:proof_prop_1}.

\subsection{Lindblad equation for a sub-spin-network}

Following the idea of the adiabatic elimination theory, an effective model is introduced to describe dynamics of sub-spin-networks. The main difficulty for unifying spin-foam theory and open quantum system theory concerns the problem of time. There is no explicit time in spinfoam theory. However, a spin-network describes a state of space, and the transition amplitude $W(\psi_{out},\psi_{in})$ includes a notion of time \cite{rovelli_covariant_2014}, since $|W(\psi_{out},\psi_{in})|^2$ gives the probability to observe $\psi_{out}$ knowing $\psi_{in}$. A change of state implies a notion of transformation, and thus, a time effect. Hence, we can formally define:
\begin{equation}
W(\psi_{out}(t'),\psi_{in}(t)) = \bra{\psi_{out}} e^{ -\ii (t'-t) \Omega } \ket{\psi_{in}}.
\end{equation}
The Hamiltonian formulation of loop quantum gravity dynamics still suffers from unsolved problems, and it is not totally clear if spinfoam transition amplitudes are always unitary.
For convenience, we assimilate the unitary part of the Lindblad equation to the spinfoam transition amplitude. In case of a non-unitary spinfoam, we can proceed in a similar way and define: $W(\psi_{out}(t'),\psi_{in}(t)) = \text{Tr}[\rho_{out}^\dagger e^{ (t'-t) \Mc L_0 } \rho_{in}]$. The adiabatic elimination presented in section \ref{sec:Adiabatic elimination for open quantum systems} works with both unitary or non-unitary systems.\textit{ Then, we keep this definition as a formal analogy, a useful trick for calculations.}

Next, we consider that the interaction with the environment is made only at $t= Cst$ times. Such interaction is formalized with operators $O : \Mc H_\Gamma \rightarrow \Mc H_\Gamma$. In this context, we define:
\begin{definition}
\label{def:master_equation_full_system}
Let $\rho$ be a density operator on $\Mc H_\Gamma$, and $R_n$ a family of operators on $\Mc H_\Gamma$. The master equation governing the evolution of $\rho$ is defined by:
\[
\frac{d\rho}{dt} = -\ii [\Omega,\rho] + \sum_{k>0} \delta (t-t_k) \sum_{n} D_{R_n} [\rho]. 
\]
\end{definition}

For simplicity, $R_n$ is assumed constant over time, but this  hypothesis can be relaxed in order to describe non-Markovian dynamics. The integration of this equation gives:
\begin{equation}
\rho(t_{k_f}) = \left[ \mathbb{T} \prod_{k=0}^{k_f} e^{ D_{R_n}} e^{- \ii (t_k - t_{k-1}) \text{Adj}_\Omega} \right] \rho(0),
\label{eq:split_operator_lindblad}
\end{equation}
with $\text{Adj}_\Omega [\rho] \equiv [\Omega,\rho]$, and consequently,  $ e^{- \ii (t_k - t_{k-1}) \text{Adj}_\Omega}\rho = W \rho W ^\dagger$. $\mathbb{T}$ denotes the time ordering of the product (a similar equation can be written for non-unitary spinfoams).

Let us have a moment of reflection on the nature of damping operators $R_n$. They act on a full spin-networks (the entire space), and they can be interpreted as the action of matter fields (in a large sense) on the gravitational field. The precise definition of these operators depends on many factors: the nature of the interaction between fields, the spatial distribution of energy, etc. Obviously, this is a very challenging task to present a well-defined and rigorous derivation of these operators, and such a complete description can be hoped only with a complete theory of quantum gravity. Maybe this step will be completed with novel results based on coarse-graining and spinfoam renormalization \cite{oeckl_renormalization_2006,banburski_pachner_2015,chen_bulk_2016,dittrich_coarse_2016,feller_surface_2017,delcamp_towards_2017}. 

The problem is avoided by means of a phenomenological model. Since we are not interested in a process influencing the entire universe, but to a process restricted to a very small portion of space, the process is modeled  with an effective Lindblad equation. This equation is destined to provide dynamics of a density operator defined on sub-spin-networks. This procedure has the disadvantage that there is no explicit mechanism that allows us to deduce operators $R_n$, but they can be easily defined, and they offer the possibility to test the theory directly. The detrimental impact of this approach is limited because the derivation of a Lindblad equation from first principles requires many assumptions dispersed at each calculation steps. Here, the assumptions are made directly during the definition of the effective model.

To proceed further, we must justify the assimilation of a set of sub-spin-networks to a space $\Mc H_0$. We notice that there exist operators which leave a sub-spin-network invariant: $O(\ket{\psi_{\Gamma}} \sqcup \ket{\psi_{\bar \Gamma '}}) = \ket{\psi_{\Gamma} }\sqcup \ket{\phi{\bar \Gamma '}}$. The operator $P_{\Mc H_{\Gamma'}\otimes \Mc H_{\bar \Gamma'} \rightarrow \Mc H_{\Gamma}} (\Id \otimes D)$, with $D$ diagonal, is a simple example. Hence, we can construct a sub-Hilbert space $\Mc H_0 \subset \Hilbert_{\Gamma}$ and the corresponding space $\Mc D_0$, such that these subspaces are associated with a set of sub-spin-networks. To construct an evolution operator with space of steady state $\Mc D_0$, we can proceed as follows: first, we choose a basis $\{ \ket{v} \}$ in the orthogonal supplement of $\Mc H_0$. Then, we define:  $P^{(m)}_k =\sum_v \ket{m} \sqcup\ket{k}\bra{v}$, and the Lindblad super-operator $\sum_k D_{P^{(m)}_k}$. Notice that $P^{(m)}_k$ maps vectors outside $\Mc H_0$ into $\Mc H_0$. Therefore, $U = \lim_{\lambda \rightarrow \infty} \exp(\lambda \sum_k \kappa_k D_{P^{(m)}_k})$, is an operator that maps a density operator $\rho \in \Mc D$ into $\Mc D_0$. 
\begin{hyp}
\label{hyp1}
We restrict this study to a class of problem where damping operators of the system without perturbation ($\epsilon = 0$) are of the form:
\[
U = \lim_{\lambda \rightarrow \infty} \exp\left(\lambda \sum_k \kappa_k D_{P^{(m)}_k}\right) ~;~ P^{(m)}_k =\sum_{v \not \in \Mc H_0} \ket{m}\sqcup\ket{k}\bra{v}~;~\kappa_k\in\setR.
\]
\end{hyp}
 There are two dynamical mechanisms that can be emphasized:
\begin{itemize}
\item If $\Mc H_0$ is an invariant subspace of $W$, the system evolves freely in this subspace and the non-unitary operator $U_0$ do not affect dynamics.
\item Any state outside $\Mc H_0$ is mapped into $\Mc H_0$ by $U$. The precise definition of this mapping is given by operators $P^{(m)}_k$.
\item If $W$ do not preserve $\Mc H_0$, we observe simultaneously the two effects.
\end{itemize}

From these observations, we construct an effective Lindblad equation, as in proposition~\ref{prop:adiabatic_elimination}. By definition, the coefficient $\kappa_{nm}$ is the probability to measure the state $n$ under the assumption that the state $m$ is initially measured. This quantity is therefore directly related to the spinfoam transition amplitude and it can be determined with its explicit computation. From the discussion here above, there are two contributions in this probability: the contribution from states inside $\Mc H_0$, and a contribution from states outside $\Mc H_0$.
\begin{hyp}
\label{hyp2}
In this paper, we restrict the study to the case $\kappa_{nm} = |W(n,m)|^2/\sum_n |W(n,m)|^2$.
\end{hyp}

This hypothesis is motivated by the following observation, for a conventional open quantum system, the probabilities to find a specific state of the reduced system is given by diagonal elements of the density matrix \cite{breuer_theory_2007} $\rho_A(t) = \text{Tr}_B [ \rho(t) ]$,  with $\text{Tr}_B$, the partial trace over the bath degrees of freedom. Explicitly, the probability to find the state $\ket{n}$, is given by $\bra{n} \rho_A(t) \ket{n}$. If we assume that the initial state is a pure state of the form $\ket{m} \otimes \ket{\psi_B}$, the probability to find $\ket{n}$ with the prior knowledge of the state $\ket{m}$ is $\bra{n} \rho_A(t) \ket{n} =\bra{n} \text{Tr}_B [ \rho(t) ] \ket{n} = \sum_{k,\mu} \bra{n,k} M_\mu \ket{m , \psi_B}\bra{m , \psi_B} M_\mu^\dagger \ket{n,k} = \sum_k |\bra{n,k}M_\mu \ket{m, \psi_B}|^2 $. The hypothesis~\ref{hyp2} can be thought as a special realization of this principle with sub-spin-networks. The normalization factor is introduced because spinfoam transition amplitudes are not necessarily normalized in calculations.

Finally, we have to compute  $W(\psi_{out},\psi_{in})$, with $\ket{\psi_{out}}, \ket{\psi_{in}}$, two states in $\Mc H_0$. Contrary to the standard Regge calculus in classical general relativity \cite{gentle_regge_2002}, we cannot work with a huge lattice, due to the enormous numerical cost required to compute a spinfoam transition amplitude. To perform the calculation, a last approximation is used:
\begin{hyp}
\label{hyp:3}
Consider the transition amplitude $W(\psi_{out},\psi_{in}) = \sum_{F: \psi_{in}\rightarrow\psi_{out}} \omega(F) Z(F)$. We conjecture that there exists a foam $F'$ on a small complex $\tilde \Gamma '$ (small number of vertices), and a spin-network $\ket{\psi^{reduced}}=\ket{\psi_{in}^{(reduced)}}\sqcup \ket{\psi_{bath}}\sqcup\ket{\psi_{out}^{(reduced)}}$ such that:
\[
W(\psi_{out},\psi_{in}) \approx Z(F') ~~;~~ F': \emptyset \rightarrow \ket{\psi^{reduced}}
\]
\end{hyp}
The quality of the approximation relies on the specific choice of $\psi_{bath}$, and on the structure of the 2-complex of $F'$. This approximation is justified with the following argument:
even if a spinfoam has a few vertices, it has a very large number of degrees of freedom, and identical input/output relations shall be obtained with different systems as soon as the parameters of the boundary spin-network are carefully adjusted.
We refer to Appendix \ref{sec:numerical_investigation_hyp_3} for a short numerical analysis of this hypothesis. 

\begin{figure}
\begin{center}
\includegraphics[width=\textwidth]{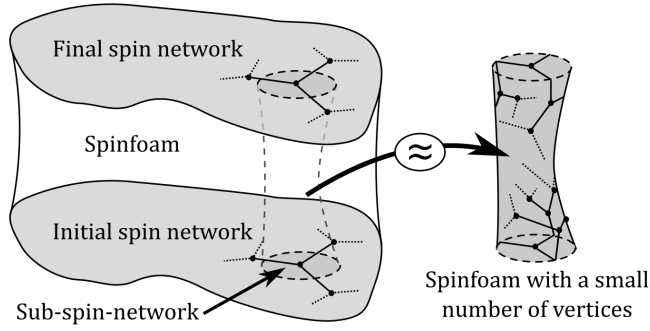}
\end{center}
\caption{Summary of the procedure used for approximated calculations of spinfoam transition amplitude.}
\label{fig2}
\end{figure}

The reduced foam $F'$ describes dynamics of space-time in a localized area. This evolution is constrained by the state on the boundary, given by $\psi_{bath}$. This "bath" degrees of freedom capture the constrains induced by the initial spinfoam and the relaxation process into $\Mc H_0$. The idea at the origin of the approximation can also be described as follows: the goal consists in  extracting the part of the spinfoam $F$ that influences the most of the state transition amplitude, and to reconstruct a spin-network on the boundary of the extracted foam to recover (approximately) the initial result. The procedure is illustrated  in figure \ref{fig2}.

Hypotheses \ref{hyp1}-\ref{hyp:3} give us a framework for which, it is only necessary to specify a Hilbert space of sub-spin-networks related to $\Mc H_0$, a 2-complex $\tilde \Gamma '$, and a state $\psi_{bath}$. These entities are assumed to describe dynamics of a more fundamental system. In the following, they are considered as inputs in the theory. We will not refer to the initial system anymore. In order to avoid confusion between the reduced system and the initial one, we introduce an abstract Hilbert space $\Mc H_r$ isomorphic to $\Mc H_0$. From this Hilbert space, we define a set of density operators $\Mc D_r$. The first order expansion of the reduced dynamics is given by the equation:
\begin{equation}
\rho(t_{k+1}) = \exp \left( g \sum_{m,n} \kappa_{nm} D_{R_{nm}} \right) \rho(t_k),
\label{eq:linblad_operator_reduced_system}
\end{equation}
with $g = \epsilon (t_{k+1}-t_k) ~\forall k$, $\rho(t_k) \in \Mc D_r$, $R_{nm}: \Mc H_r \rightarrow \Mc H_r$ such that $R_{nm} = \ket{n}\bra{m}$, $\kappa_{nm} = |W(n,m)|^2/(\sum_m |W(n,m)|^2)$, and $|W(n,m)|^2 = Z(F: \emptyset \rightarrow \ket{\psi_{m}^{(reduced)}}\sqcup \ket{\psi_{bath}}\sqcup\ket{\psi_{n}^{(reduced)}})$. Equation~\eqref{eq:linblad_operator_reduced_system} is the formal integration of the system of the proposition \ref{prop:adiabatic_elimination}. The dimensionless constant $g$ captures the coupling constant and the time scale of the non-unitary process of the weakly coupled subsystem. Notice that equation~\ref{eq:linblad_operator_reduced_system} can also be derived from a stroboscopic dynamic and a first order expansion in $g$ of the exponential.

Here, time is introduced in a special way. It is defined by coherent dynamics, given by $W(\psi_{out}(t'),\psi_{in}(t))$, but it also depends on the relaxation process. The definition of time makes sense only because there is an underlying system for which we have an implicit reference. Time is discrete because the interaction with the environment happens only at specific time slices, but the process has a continuous degrees of freedom due to the coupling constant $\epsilon$.

The time interval $[t_k,t_{k+1}]$ can be inferred from the reduced boundary state (see \cite{rovelli_covariant_2014} and \cite{christodoulou_characteristic_2018,christodoulou_planck_2016,christodoulou_transition_2017,dambrosio_end_2020} for applications). However, the only relevant parameter in equation~\ref{eq:linblad_operator_reduced_system} is $g$. Then, it is not necessary to derive the time step explicitly. We can keep $\epsilon$ and $(t_{k+1}-t_k)$ unspecified.

Based on the mathematical structure of equation~\eqref{eq:linblad_operator_reduced_system}, we can make an analogy with an ensemble of qubits coupled with a strongly dissipative cavity \cite{breuer_theory_2007}. The states of the qubits are similar to the sub-spin-networks, and the cavity plays the role of the gravitational bath. The cavity is described by a density of states \cite{garraway_nonperturbative_1997}. For the gravitational system, the bath is described by a quantum state on the complementary of the input/output sub-spin-networks. 

\section{Decoherence of the gravitational field}
\label{sec:decoherence of the gravitational field}

In this section, the theory developed previously is illustrated by means of examples. In particular, the decoherence of the gravitational field is studied. First, a qualitative study is made using a two-level system (dim$(\Mc H_r)=2$), and then, numerical results are presented for systems of higher dimensions (dim$(\Mc H_r)\geq 4$).

\subsection{Analytic investigation}
\label{sec:analytic investigation}

To compute analytically the dynamics of a two-level system, i.e. a reduced Hilbert space composed of two states labeled $1$ and $2$, we consider a system for which "\textit{in}" and "\textit{out}" states are independent variables such that $W_{nm} \propto W_n^*W_m$.

A straightforward application of equation \eqref{eq:linblad_operator_reduced_system} gives:
\begin{equation}
\begin{split}
\kappa_{11}& \propto \frac{|W_1|^4 }{|W_1|^{2} + |W_1 W_2^*|^{2}} \\
\kappa_{12}& \propto \frac{|W_1 W_2^*|^2 }{|W_1|^{4} + |W_1 W_2^*|^{2}}\\
\kappa_{21}& \propto \frac{|W_2 W_1^*|^2}{|W_2|^{4} + |W_2 W_1^*|^{2}}\\
\kappa_{22}& \propto \frac{|W_2|^2 }{|W_2|^{4} + |W_2 W_1^*|^{2}}
\end{split}
\label{eq:def_RNM_operator_2lvl_syst}
\end{equation}
Solutions of $ \sum_{n,m} \kappa_{n,m}\Mc D_{R_{nm}} [\rho ] = 0$ gives us the steady state:
\begin{align}
\label{eq:steady_state_2lvl_syst}
 \rho_{11}& = \frac{\kappa_{12}}{\kappa_{12} + \kappa_{21}}\\
 \rho_{22}& = 1-  \rho_{11} \\
 \rho_{12}& =  \rho_{21} =0
\end{align}

\begin{figure}
    \includegraphics[width=\textwidth]{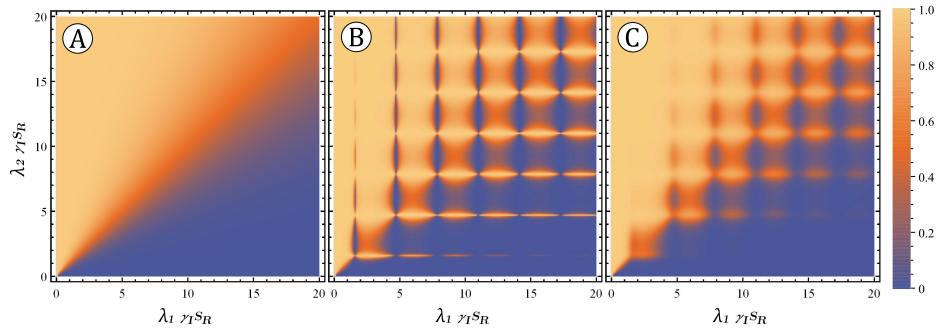}
    \caption{Steady state of $ \rho_{11}$ as a function for several values of $\alpha$. (A) $\alpha = 0$, (B) $\alpha = 1$, (C) $\alpha = 2$.
    } \label{fig:steady_state_j1_j2}
\end{figure}
This state is characteristic of a mixed state. It is a well-known result in cavity-QED, when a single qubit is coupled to a dissipative environment.

Calculations are pushed further with the assumption that $W_i$ is the EPRL transition amplitude element of the boundary graph of a single vertex (5 nodes, each one connected with all others). This transition amplitude has been studied intensively during the last year~\cite{barrett_lorentzian_2010,magliaro_regge_2013,han_asymptotics_2013,rovelli_covariant_2014,dona_numerical_2019}. We can show that the amplitude is of the form : $W \propto \prod_f  (2 j_f +1) \prod_v A_v (j_f,\vec n_f)$, where $A_v(j_f, \vec n_f)$ is the vertex amplitude, a quantity that can be expressed as a function of coherent states variables $j_f$ and vectors $\vec n_f$.In the large spin limit \cite{barrett_lorentzian_2010}, we have:
\begin{equation}
A_v = \frac{(-1)^{\chi+M}}{\lambda^{12}} e^{\ii \lambda \Phi_c} \left( N_+ e^{\ii \lambda \gamma_I S_R} + N_- e^{-\ii \lambda \gamma_I S_R} \right) + o(\lambda^{-13})
\end{equation}
with, $\chi + M \in \setN/2$, $\Phi_c \in \setR$, $N_+,N_- \in \setC$, and $\lambda$ is a scaling parameter of the spins $j_l$ that allows us to take the large spin limit. $S_R$ is the Regge action of the 4-simplex. We refer the literature for a detailed description of each parameter \cite{dona_numerical_2019,barrett_lorentzian_2010}. There are 10 links in the boundary graph, hence, 
$
W(\lambda) \propto \lambda^{10}  \prod_{l=1}^{10} 2 j_l ) A_v (\lambda
$
. Notice that the area is proportional to $\lambda$: $A_l =8 \pi \gamma_I \sqrt{\lambda j_l (\lambda j_l +1)} \simeq 8 \pi \gamma_I \lambda j_l$ (see equation~\eqref{eq:face_area}).  Using, the approximated expression of $W$, two values of $\lambda$ are used to parametrize the reduced system, and inserting \eqref{eq:def_RNM_operator_2lvl_syst} into \eqref{eq:steady_state_2lvl_syst}, we obtain:
\begin{equation}
 \rho_{11}= \frac{\lambda_1^4 f(\lambda_2)}{\lambda_1^4 f(\lambda_2)+\lambda_2^4 f(\lambda_1)} ~;~f(\lambda) = \vert e^{2 \ii \lambda \gamma_I S_R} + \alpha \vert \vert N_+ \vert,
 \label{eq:steady_state_j1_j2}
\end{equation}
with $\alpha = N_+/N_-$. This expression is plotted in figure \ref{fig:steady_state_j1_j2} for three values of $\alpha$. We observe that in the case $\lambda_1 \ll \lambda_2$ we have $ \rho_{11} = 0$, or $1$ inversely. \textit{This is a relaxation effect toward a state of small area.} In the general case, we have a statistical superposition of the two states that can be described with an effective temperature. For $\alpha \neq 0$, we observe a complex structure in the steady state with switchings of the population between $\rho_{11}$ and $\rho_{22}$. Non-trivial steady states shall be observed if many states are taken into account in the reduced dynamics. 

\subsection{Numerical simulation}
\label{sec_numerical_simulation}

We continue to illustrate the theory with a few numerical simulations. Three different reduced spinfoams  of 2-vertices with different types of structures, and different reduced boundary states are considered. For now, spinfoams with more than 2 vertices or more complex structures are out of reach of numerical calculations. 

For simplicity, we work with toy models of spinfoams without internal faces, which means that all 3-simplices of the foam belong to the boundary. Only a few degrees of freedom are allowed to change. These systems can be viewed as low order approximations of spinfoams for which  graphs with internal faces have been neglected. Systems are defined as follows:

\begin{figure}[h]
a)
\begin{center}
\includegraphics[width=\textwidth]{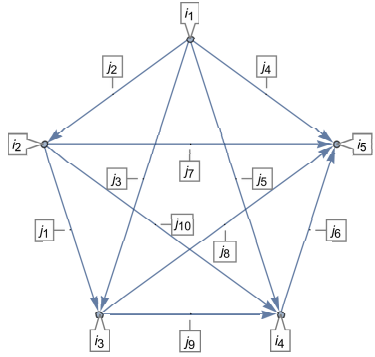}
\end{center}
b)
\begin{center}
\includegraphics[width=\textwidth]{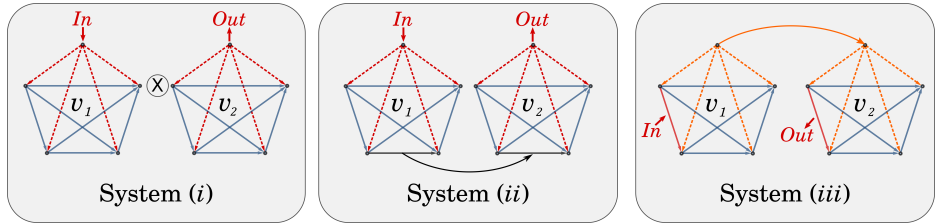}
\end{center}
\caption{
a) Boundary graph of a single vertex.\
b) Boundary graphs of the spinfoams considered in section~\ref{sec_numerical_simulation}. Links and nodes shared by two vertices are represented by a different line-style with an arrow connecting the vertices.}
\label{fig:triangulation}
\end{figure}

\begin{enumerate}
\item A spinfoam composed of two vertices without commune faces, like in section\ref{sec:analytic investigation}.
States $\ket{\psi^{(reduced)}_{in/out}}$ are given by an intrinsic coherent state \cite{livine_new_2007,rovelli_covariant_2014} that describes a regular tetrahedron of spin label $j \in \{1/2,1,3/2,2\}$. The canonical basis of $\Mc H_r$ is labeled by these spin numbers. By convention "$in$" and "$out$" tetrahedrons are placed at the node "1" of the corresponding vertices (see figure \ref{fig:triangulation}). The bath wave function is chosen to be:
\begin{equation}
\ket{\psi_{bath}} = \sum_{j_l,\i_n} \Mc A_l (j_l) e^{ -\frac{(j_l-j_0)^2}{\sqrt{j_0}}}
\ket{\{\i_n\},\{ j_l\}},
\label{eq:psi_bath_1}
\end{equation}
with $j_0 = j_{in}$ if the corresponding link is in $v_{1}$, and $j_{out}$ otherwise. The sum runs over all links and nodes that do not belong to "$in$" and "$out$" tetrahedrons. The weights introduced in equation~\eqref{eq:psi_bath_1} are introduced to mimic a semi-classical extrinsic geometry \cite{bianchi_coherent_2010} whose mean value is given by $j_0$.
This model corresponds to disconnected manifolds. This might be an interesting approximation of two subsystems sufficiently far from each other such that their transition amplitude can be factorized.

\item A spinfoam of two vertices connected by a single link (by convention, the link "9" of the vertex, see figure~\ref{fig:triangulation}). This system can be viewed as an extension of system $(i)$ with a "minimal coupling" between vertices. States are defined similarly as system ($i$), except that we symmetrize the amplitude of the link "9" by averaging $j_9$ over the values $j_0 = j_{in}$ and $j_{out}$.

\item A spinfoam composed of two vertices connected by a node and its links (by convention, we use the node "1" and the links "2,3,4,5"). States $\ket{\psi^{(reduced)}_{in/out}}$ are given by the state of a single link $\ket{j}$, with $j\in\{1/2,1,3/2,2,5/2,3,7/2,4\}$. The canonical basis of $\Mc H_r$ is labeled by these spin numbers. By convention "$in$" and "$out$" faces are placed at the face "1" of the corresponding vertex. The bath state is:
\begin{equation}
\ket{\psi_{bath}} = \frac{1}{2} \sum_{\alpha=1}^2 \sum_{(j_l^\alpha,\i_n^\alpha) \in v_\alpha}\sum_{(j_l^{\alpha +1},\i_n^{\alpha+1}) \in v_{\alpha +1} \smallsetminus v_\alpha}  \delta_{j_l^\alpha}^{j_{in}}\delta_{\i_n^\alpha}^{\max(\i_n^\alpha)}   \delta_{j_l^{\alpha +1}}^{j_{out}}\delta_{\i_n^{\alpha+1}}^{\max(\i_n^{\alpha+1})} \ket{\{\i_n\},\{ j_l\}},
\label{eq:state_bath_case_3}
\end{equation}
where $\max(\i_n^a)$, is the maximum value allowed by $\i_n^a$. The sum runs over all links that do not belong to the boundary, and the summation variable is defined such that $\alpha +1 = 1$ if $\alpha = 2$. 
Contrary to systems ($i$) and ($ii$), which have some semi-classical properties, this system is highly non-classical (by analogy with QED, this system can be compared to a $n$-particle state, while other systems are closer to semi-classical systems).
\end{enumerate}

These systems are physically interesting for studies of small portions of the gravitational field strongly constrained by the surrounding field (since all 3-simplices are on the boundary, which is fixed by the spinfoam boundary state).

We remark that these systems can be viewed as prototypes of spin-networks for studding black holes \cite{perez_black_2017,christodoulou_planck_2016,christodoulou_characteristic_2018,dambrosio_end_2020}, for which the gravitational bath depends on the geometry of the boundary state. Indeed, the dynamic horizon of a black hole is related to its mass $M$. If the horizon is defined using a collection of faces of spin $j_0$ we obtain a mapping $M(j_0)$. Each small volume of space surrounding the horizon can be chosen flat, but with an extrinsic geometry $K(j_0)$ capturing the gravitational effect of the mass $M(j_0)$. The extrinsic geometry is here captured by the weight $e^{-(j-j_0)^2/\sqrt{j_0}}$ in equation \eqref{eq:psi_bath_1} \cite{bianchi_coherent_2010}. However, the modeling of a black hole with a reduced spinfoam is subject to many subtle points that we do not pretend to solve in this paper.

Transition amplitude coefficients are determined numerically using the \emph{SL2Cfoam -v1} library \cite{dona_numerical_2018}. The preliminary computation of EPRL-Hashtables (tables of data required to evaluate a vertex amplitude) has taken approximately two weeks on a computer using 32 cores and 128 GB of RAM. Once this step completed, the evaluation of transition amplitude elements varies between a few seconds to several hours, depending on the number of graphs taken into account. Due to the Gaussian cutoff of the internal face amplitudes (cases ($i$) and ($ii$)), a maximum of $j_f = 4$ has been  used in calculations. Additionally, the cutoff parameter of \emph{SL2Cfoam-v1} has been set to $\Delta = 4$. According to \cite{dona_numerical_2018}, this value of $\Delta$ gives a result with an error of $\simeq$ 10\% . This choice is a good compromise between accuracy and computation speed.  The dimension of the system (number of vertices) is limited by the memory of the computer. More elaborate codes (machine learning to determine which graphs are the most relevant, dynamic management of loaded HashTables,...) shall reduce the computation cost.

\begin{figure}[h]
\includegraphics[width=\textwidth]{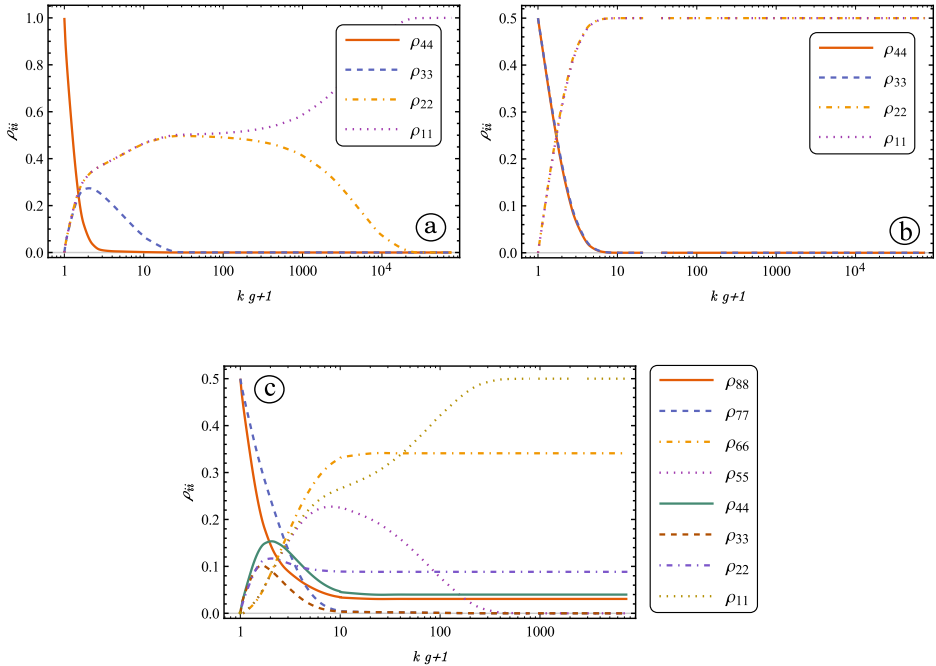}
\caption{Evolution of diagonal density matrix elements $\rho_{ii}$ as a function of the iteration number $n$ for the 3 reduced systems introduced in sections \ref{sec_numerical_simulation}. Sub-figures a,b,c correspond respectively to systems ($i$), ($ii$), ($iii$). Initial states are pure state respectively given by $\ket{2}$, $(\ket{2} + \ket{3/2})/\sqrt{2}$, and $(\ket{4} + \ket{7/2})/\sqrt{2}$.}
\label{fig:grav_decoherence}
\end{figure}

For each system, matrix elements $W_{nm} = W(j_n,j_m)$ are computed and probability damping coefficients $\kappa_{nm}$ and operators $R_{nm}$ are deduced from the hypothesis \ref{hyp2} and proposition~\ref{prop:adiabatic_elimination}. Once an initial state is chosen, the reduced dynamic is computed straightforwardly using equation~\eqref{eq:linblad_operator_reduced_system}. As illustrative examples, the time evolution of density matrix diagonal elements of the three reduced systems are plotted in figure \ref{fig:grav_decoherence}.

First of all, it should be underlined that systems ($ii$) and ($iii$) have fundamentally different structures of interaction between $in$ and $out$ states than system ($i$), but, system ($i$) and $(iii)$ have qualitatively similar dynamics. This similitude suggests that even if the gravitational bath in a highly non-classical state or a semi-classical one, the resulting dynamics are comparable, and system ($i$) can approximate more complex spinfoams. This point is also discussed in Appendix \ref{sec:numerical_investigation_hyp_3}.

The second observation joins the result of section \ref{sec:analytic investigation}: we observe a decoherence toward a ground state of smaller area. However, we can also verify (not shown in the graphs), that the opposite is also possible if the steady state has an average area  larger than the initial state. Moreover, in all cases, the time scales of the different transitions are sufficiently different (several orders of magnitudes) so that only 2 or 3 states evolve simultaneously. This point can be crucial for further studies. Indeed, it may not be necessary to work with a huge spinfoam, but to restrict the gravitational system to a few effective boundary states, like in reference \cite{rovelli_small_2018}.

\begin{figure}[h]\sidecaption
\includegraphics[width=8cm]{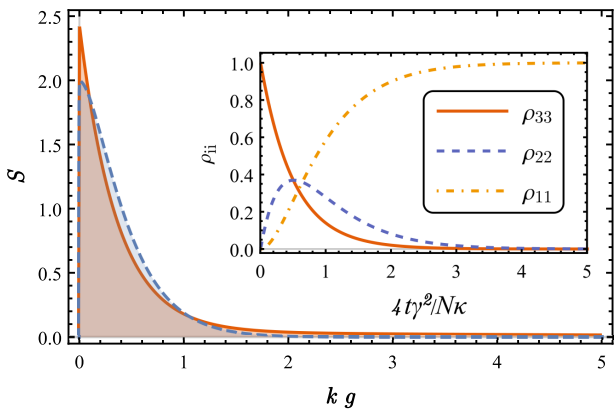}
\caption{Comparison between the energy released by the relaxation of system ($i$)(orange solid line), with the decay of $N=2$ qubits coupled at resonance with a cavity (blue dashed line). Curves are normalized by their integral over time. The cavity damping rate is $\kappa/\gamma = 40$, with $\gamma$ the coupling strength \cite{haroche_exploring_2006,louisell_quantum_1973}. The inset shows the evolution of qubit state's population in the Dicke basis $\ket{1} = \ket{1,-1}$, $\ket{2}=\ket{1,0}$, $\ket{3} = \ket{1,1}$. Time is rescaled by the damping rate. These dynamics must be compared with the ones plotted in figure \ref{fig:grav_decoherence} (a).}
\label{fig:radiation_energy}
\end{figure}

Dynamics plotted in figure~\ref{fig:grav_decoherence} are similar to dissipative dynamics of atomic systems \cite{gross_superradiance:_1982,santos_master_2014}. Following the standard ideas in cavity QED, we can compute the energy of the radiation produced by the relaxation process. For that purpose, we use the fact that the energy of a face is proportional to its area \cite{bianchi_entropy_2012}, and we define the operator acting on the reduced system: $\hat E = \lambda \sum_{j_1,j_2} \delta_{j_1,j_2} \sqrt{j_1(j_1+1)}\ket{j_1}\bra{j_2}$, with $\lambda$ a proportionality constant. Then, we deduce the energy released as a function of time: $S(t_k) = - \partial_{t_k}\langle \hat E(t_k)\rangle $, with $\langle \hat E(t_k)\rangle = \text{Tr}[\rho(t_k) \hat E ]= \epsilon  \text{Tr} [\Mc L \hat E]$, with $\Mc L$ the lindblad super-operator  of equation~\eqref{eq:linblad_operator_reduced_system}. The energy loss during the relaxation of system ($i$) is plotted in figure \ref{fig:radiation_energy}. We notice the strong similitude with the energy loss of an ensemble of qubits coupled with a bad cavity. In that case, the energy released by the qubits is given by $\langle \hat S_z \rangle (t)$  \cite{haroche_exploring_2006,louisell_quantum_1973}. The best match between the curves is found for a system of two qubits at resonance, for which 3 energy levels are coupled together. This is in agreement with the  dynamics of the gravitational system: over the time interval of comparison  the gravitational dynamic is reduced to 3 states (see figure \ref{fig:grav_decoherence}).

Many properties of quantum systems are derived from thermodynamic quantities, such as entropy or temperature. Among the possible applications to quantum gravity, the thermal time hypothesis \cite{connes_von_1994,rovelli_statistical_1993,rovelli_quantum_2007} is an interesting proposal of definition for the time flow. In the rest of this section, we explore to which extant thermal time is a relevant concept of time for reduced quantum gravitational systems.

The origin of the time variable is discussed in section~\ref{sec:effective_model}. It is argued that time is defined with the assumption of a larger system for which the reduced system is defined. However, it could be interesting to derive the notion of time with only the information available from the reduced gravitational system, and nothing else. The thermal time hypothesis  is a proposal that goes toward this direction. Explicitly, the thermal time flow is realized as a $\sigma$-weakly continuous one-parameter group of $^*$-automorphisms on the observable algebra. Given a state $\rho$, we define the thermal Hamiltonian $K=-\text{ln}(\rho)$, and the thermal time $s$, such that they generate a transformation of the form: $e^{\ii s K}$ \cite{connes_von_1994,rovelli_covariant_2014,paetz_analysis_2010}. Thermal time has interesting applications in the context of states at thermal equilibrium. In particular, it is possible to show that temperature is "the speed of time" in a static space-time \cite{menicucci_clocks_2011}. Open quantum systems considered in this paper are out of equilibrium, then, we discuss the extent to which the thermal time hypothesis can be applied. 

\begin{figure}[h]\sidecaption
\includegraphics[width=8cm]{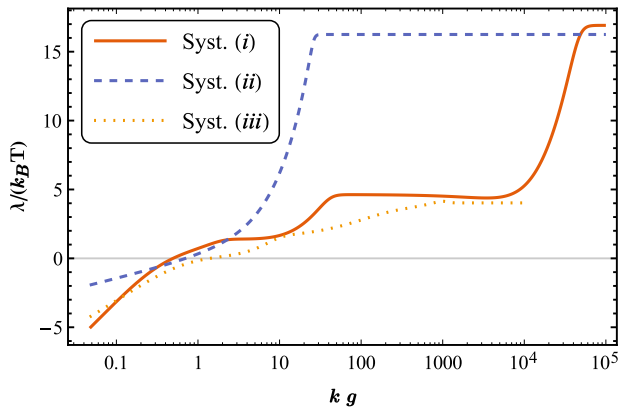}
\caption{Evolution of the spectral temperature for systems $(i)$ to $(ii)$.}
\label{fig:thermal_time}
\end{figure}

First we explore the idea that temperature is the speed of time. In the context of an open quantum system, it is necessary to generalize the definition of temperature. Here, we employ the spectral temperature \cite{gemmer_quantum_2009}:
\begin{equation}
\frac{1}{k_B T} = -\left(1-\frac{\rho_{11}+\rho_{NN}}{2}\right)^{-1} \sum_{i=2}^N \left( \frac{\rho_{ii}+\rho_{i-1,i-1}}{2}\right) \frac{\ln(\rho_{ii}/\rho_{i-1,i-1})}{E_{i} - E_{i-1}}
\end{equation}
with $N = \text{dim}( \Mc H_r)$, $W_i = \rho_{i/2,i/2}$ and $E_{i}$ is the energy of the level $i$. The evolution of the spectral temperature for the three systems is plotted in figure \ref{fig:thermal_time}.  

Two main observations are made. First, the temperature is negative for $k g \simeq 0$, and then become positive when the state approaches a thermal state. We can check that the steady state is a thermal state with the corresponding temperature. Secondly, temperature does not evolve monotonically, as around $k g \simeq 100$ for system ($i$), or $k g \simeq 1000$ for system ($ii$). From these observations, we conclude that the temperature is not a parameter that allows us to characterize the "speed" of the time flow (at least, for the systems presented here).

Moreover, we notice that a thermal time $s_k$ can be defined for any $\rho(t_k)$, and these $s_k$ are not necessarily related: $\forall~ \rho(t_k), ~\exists~ s_k $such that $ e^{\ii s_k \ln \rho(t_k)}$ is a thermal flow. Here, the time of the Lindblad equation generate a family of thermal time, but not the reciprocally. The flow defined by $ e^{\ii s_k \ln \rho(t_k)}$ is different from  the one defined by $e^{k g \Mc L}$. Therefore, the thermal time is an interesting parameter to describe dynamics under the assumption that the system of reference is stationary, but it cannot be used to describe out of equilibrium dynamics of the gravitational field. These results are in agreement with the analysis given in \cite{paetz_analysis_2010}.

\section{Conclusion}
\label{sec:conclusion}

This paper is devoted to the definition of an open quantum system theory of a portion of space-time in interaction with an environment. Space-time quantization is made in the framework of loop quantum gravity. This open quantum system theory is based on two main ingredients. The first one is the extraction of dynamic properties of the gravitational field with spinfoam calculations, and the second is the definition of an effective model by following concepts of adiabatic elimination of Lindblad equations. As a result, we obtain a Markovian master equation with damping operators that depends on spinfoam transition amplitudes. These quantities can be computed using a brute-force numerical computation. Several physical aspects of this theory are investigated by means of test examples. In particular, it is shown that relaxation toward a state of small area is possible. This is analogous to the Purcell effect in the bad cavity regime. All dynamics studied in this paper have very different time scales that involves generally two or three different states. This suggests that effective dynamics of the quantum gravitational field can be modeled with a few relevant quantum states. Finally, a possible relation with thermal time and the Markovian process of effective dynamics has been studied. It has been illustrated that temperature cannot be assimilated, in general, to the speed of time, and that the time parameter generates a family of thermal time parameters.

Open-spinfoams presented in this paper are interesting to underline some properties that we can reobserve with other models, but they could have limited applications.
It can be interesting to extend this formalism with matter field that interacts coherently with the reduced system, and to explore more realistic physical situations in black hole physics \cite{christodoulou_planck_2016,christodoulou_transition_2017,christodoulou_planck_2016,dambrosio_end_2020} and cosmology. 
Further advances in coarse-graining and spinfoam renormalization could be useful for the design of more interesting reduced systems.

\begin{acknowledgements}
Thanks to Giorgio Sarno for helpful discussions about the SL2Cfoam code, and to David Viennot for advised comments. Simulations have been executed on computers from the Utinam Institute of the Université de Franche-Comté, supported by the Région de Franche-Comté and Institut des Sciences de l'Univers (INSU).
\end{acknowledgements}

%
\section*{Conflict of interest}
 The author declares that he has no conflict of interest.

\appendix

\section{Some mathematical definitions}
\label{sec:math_def}

This appendix is devoted to some mathematical definitions, which are omitted in section~\ref{sec1} for conciseness.  Further technical details on this subject can be found in Refs.~\cite{baez_introduction_1999,rovelli_quantum_2007,rovelli_covariant_2014,livine_short_2011}. 

\begin{definition}
A spin-network $\psi$ is a triple $(\Gamma,\chi, \imath $)consisting of:
\begin{enumerate}
\item a 1-dimensional oriented complex $\Gamma$, represented by a graph whose vertices are points of the space-time manifold and edges are paths connecting the points.
\item a labeling $\chi$ of each edge $e$ by a unitary irreducible representation $\chi_e$ of $\mathfrak{H}$, where $\mathfrak H \subseteq \mathfrak G$ is a subgroup of the gauge group of the theory. Elements of $\mathfrak H$ associated with edges encode the structure of space at the classical level.
\item a labeling $\imath$  of each vertex $v$ by an intertwiner 
\[ \imath_v:\chi_{e_1} \otimes \cdot\cdot\cdot \otimes \chi_{e_n} \rightarrow \chi_{e'_1} \otimes \cdot\cdot\cdot \otimes \chi_{e'm}\]
where $e_1,...,e_n$ are the edges incoming to $v$ and $e'_1,...,e'_m$ are the edges outgoing from $v$.
\end{enumerate}

\end{definition}

It is usual to consider $\mathfrak G = SL(2,\setC)$ and $\mathfrak H = SU(2)$ in 3+1D gavity, or $\mathfrak G = \mathfrak H = SU(2)$ in 3D Euclidan gravity \cite{rovelli_covariant_2014}. For reasons that will become clearer below, it is preferred to substitute the names \textit{vertex} and \textit{edge} by \textit{node} and \textit{link}.

\begin{definition}
\label{def:spinfoam}
A spinfoam $F$ is a triple $(\tilde \Gamma,\tilde \chi,\tilde \imath $)consisting of:
\begin{enumerate}
\item a 2-dimensional oriented complex $\tilde\Gamma$.
\item a labeling $\tilde \chi$ of each face $f$ by a unitary irreducible representation $\tilde \chi_e$ of $\mathfrak{G}$.
\item a labeling $\tilde \imath$  of each vertex $e$ by an intertwiner 
\[\tilde \imath_e:\tilde \chi_{f_1} \otimes \cdot\cdot\cdot \otimes \tilde \chi_{f_n} \rightarrow \tilde \chi_{f'_1} \otimes \cdot\cdot\cdot \otimes \tilde \chi_{f'_m}\]
where $f_1,...,f_n$ are faces incoming to $e$ and $f_1',...,f_m'$ are faces outgoing from $e$.
\end{enumerate}
\end{definition}
Notice that here, vertices are D dimensional objects, edges are D-1 dimensional objects, and faces are D-2 dimensional objects.
\begin{definition}
Let $\psi$ be a spin-network $\psi=(\Gamma,\chi,\imath)$. A spinfoam $F:\emptyset \rightarrow \psi$ is a triple $(\tilde \Gamma,\tilde \chi, \tilde \i)$ verifying the definition \ref{def:spinfoam}, and:
\begin{enumerate}

\item for any link $l$ of $\Gamma$, $\chi_l = P(\tilde \chi_f)$ if $f$ is incoming to $l$, and $(\chi_l)^* = P(\tilde \chi_f)$ otherwise.
\item for any node $n$ of $\Gamma$, $\imath_n = P(\tilde \imath_e)$.
\end{enumerate}
$P$ is a map from unitary irreducible representations of $\mathfrak{G}$ to unitary irreducible representations of $\mathfrak{H}$. It is an input of the theory.
\end{definition}
This definition gives the transition amplitude from the vacuum to a given spin-network. The evolution from a spin-network to another one is given by:
\begin{definition}
A spinfoam $F: \psi \rightarrow \psi'$ is defined by $F:\emptyset \rightarrow \psi ^* \otimes \psi'$, where $^*$ denotes the dual of a spin-network.
\end{definition}
A spinfoam model is defined with a specific choice of ($\mathfrak{G},\mathfrak{H},\imath,P,Z$).

\section{Proof of proposition 1}
\label{sec:proof_prop_1}

This appendix is devoted to the proof of proposition~.\ref{prop:adiabatic_elimination}.

\begin{proof}
We provide here a simple proof with a minimum of details. Further information is given in Reference~\cite{azouit_adiabatic_2016}.

We are looking for a master equation $d_t \rho_e = \Mc L_e \rho = [\Mc L_{e,0} + \epsilon \Mc L_{e,1} + o(\epsilon^2) ]\rho_e$ that provides the evolution of the system up to an error of $\epsilon ^2$. To determine this equation,  we introduce the map:
\[
\rho(t) = K( \rho_e(t) )= \sum_{m\geq 0} \epsilon^m K_m (\rho_e(t)).
\]
Then, by definition we have:
\[
\frac{d \rho}{dt} = \Mc L_0 K(\rho_e) + \epsilon \Mc L_1 K(\rho_e) = K(\Mc L_e \rho_e).
\]
Using the expansion of $K$ in powers of $\epsilon$, we deduce that:
\begin{equation}
\Mc L_0 K_0(\rho_e) = K_0(\Mc L_{e,0} \rho_e) ~~(m=0),
\label{eq:Lindblad_recurance_order_0}
\end{equation}
and,
\begin{equation}
\Mc L_0 K_1(\rho_e) + \Mc L_1 K_0(\rho_e) = K_0(\Mc L_{e,1} \rho_e)  ~~(m=1).
\label{eq:Lindblad_recurance_order_1}
\end{equation}
Since we are interested in dynamics restricted in $\Mc D_0$, we define $K_0$ as a projector on $\Mc D_0$: $K_0 (\rho) = P_0 \rho P_0 ^\dagger$, with $ P_0 = \sum_{n=1}^{dim \Mc H_0} \ket{n}\bra{n}$, with $\{\ket{n}\}$ a basis of $\Mc H_0$. From the definition of $\Mc D_0$, we deuce that equation~\eqref{eq:Lindblad_recurance_order_0} gives $\Mc L_{e,0}=0$. We also introduce the Kraus map of a solution of the unperturbed system when $t \rightarrow \infty$: $U_0\rho(0)=\lim_{t \rightarrow \infty}e^{t\Mc L_0}\rho(0) \equiv \sum_\mu M_\mu \rho(0) M_\mu ^\dagger$, with $\{ M_\mu\}$ an ensemble of operators such that $\sum_\mu M_\mu^\dagger M_\mu = \Id_{\Mc H} \}$.

To arrive at the desired result, we apply the super-operator $U_0$ on equation~\eqref{eq:Lindblad_recurance_order_1}. Since $U_0.\Mc L_0 =0$, we have:
\begin{equation}
U_0  \Mc L_1 K_0(\rho_e) = U_0 K_0(\Mc L_{e,1} \rho_e)
\end{equation}
Moreover, $U_0$ leaves $K_0$ unchanged. Therefore,
\begin{align}
&U_0  \Mc L_1 K_0(\rho_e) = K_0(\Mc L_{e,1} \rho_e) \\
\Rightarrow ~& K_0(U_0  \Mc L_1 K_0(\rho_e)) = \Mc L_{e,1} \rho_e
\end{align}
In the second line, we have used $K_0^2 = K_0$ and the fact that $\Mc L_{e,1}$ is restricted to $\Mc H_0$. The structure of the operator $\Mc L_1$ is by assumptions:
\[
\Mc L_1 \rho_e \equiv  \sum_{n = 1}^{dim \Mc H_0} \left( P_n \rho_e P_n ^\dagger -\frac{1}{2} P_n^\dagger P_n \rho_e - \frac{1}{2} \rho_e P_n^\dagger P_n \right)~~,~~ P_n = \ket{n}\bra{n}.
\]
Then, 
\[
\Mc L_{e,1} \rho_e \equiv  \sum_{n,m,\mu} |M_{\mu,nm}|^2 \left( \ket{n}\bra{m}  \rho_e \ket{m}\bra{n} -\frac{1}{2}\ket{m}\bra{n} \ket{n}\bra{m} \rho_e - \frac{1}{2} \rho_e \ket{m}\bra{n} \ket{n}\bra{m} \right)
\]
To finish the proof, we define $\kappa_{nm} = \sum _\mu |M_{\mu,nm}|^2$.

\end{proof}

\section{Numerical investigation of hypothesis~\ref{hyp:3}}
\label{sec:numerical_investigation_hyp_3}

In this section, we present briefly the numerical investigation that leads us to the hypothesis~\ref{hyp:3}. Contrary to examples in the main text, the 3D theory with gauge group $\mathfrak{G} = SU(2)$ is used. The transition amplitude is given by~\cite{barrett_ponzano-regge_2009,rovelli_covariant_2014}
\begin{equation}
\label{eq:transition_amplitude_3D}
W_{\tilde \Gamma} (j_l) = \Mc N_{\tilde \Gamma} \sum_{j_f} \prod_f (-1)^ {j_f} (2 j_f +1) \prod_v (-1)^{\sum_{a=1}^6  j_{v,a}} \left\lbrace \begin{array}{ccc}
j_{v,1} & j_{v,2} & j_{v,3} \\ 
j_{v,4} & j_{v,5} & j_{v,6}
\end{array} \right\rbrace,
\end{equation}
with $j_l$ the spin label of the link $l$, of a boundary spin-network. $j_f$ is the spin label of the face $f$ of the spinfoam, and $v$ is a vertex of the foam. $(v,a)$ denotes the face $a$ associated with the vertex $v$, $\{... \}$ is Wigner's 6j-symbol, and $ \Mc N_{\tilde \Gamma}$ is a normalization factor.

In the computation, we consider foams with $V= 2,3$ or 4 vertices glued one-by-one with at most one edge. The 2-complex is defined without bubbles. The transition amplitude can be written as $W(\psi_{out},\psi_{in})$, with $\psi_{in}$ a spin-network formed by the boundary of the firsts vertices, and $\psi_{out}$ the spin-network formed the boundary of other vertices. We can take symmetric networks, but it is not necessary for our purpose. For each $in$ and $out$ states, we are interested in a reduced number of degrees of freedom, given by a sub-spin-network, made of a single node and its three boundary links. The reduced Hilbert space $\Mc H_r$ is defined by the formal association of its canonical basis to a set of specific sub-spin-networks. 

For example, if we consider a sub-spin-network composed of 1 node and 3 adjacent links with spin numbers $(1,1,1)$ and $(2,2,2)$, we have: $\Mc H_r = \setC^2$,  $e_1 \rightarrow (1,1,1)$ and $e_2 \rightarrow (2,2,2)$.  For the numerical calculation presented below, sub-spin-networks are generated randomly, with dim$(\Mc H_r) = 10$. 

The transition amplitude between these sub-spin-networks is calculated using a large number ($\sim 10^6$) of different spin-networks generated randomly. This defines a map $\Mc H_r \otimes \Mc H_r \rightarrow \setC$ such that $(n,m) \rightarrow W_{nm}$. The goal is to find an approximation of $W_{nm}$ with a simpler model. Here, we choose for the simplest model,  a foam made of two unconnected vertices, such that $W \propto W_{v_1}(j_l)W_{v_2}(j_l')$. With a minimization algorithm (the \textit{NMinimize} function of \textit{Mathematica}), we find a linear superposition of spin-networks such that the transition amplitudes between sub-spin-networks of the simplified system are the closest to the values given by the initial system. To quantify the difference between models, the following cost function is used: $C= \Vert \vec W_1/\Vert \vec W_1 \Vert - \vec W_2/\Vert \vec W_2 \Vert \Vert$, where $\vec W_i$ is a vector of dimension dim$(\Mc H_r)^2$ whose components are transition amplitudes $W_{nm}$ given by the model $i$. Vectors are normalized in the definition of the cost function in order to take into account the normalization factor $\Mc N_{\tilde \Gamma}$ introduced in equation~\eqref{eq:transition_amplitude_3D}.

\begin{figure}
\begin{center}
\includegraphics[width=\textwidth]{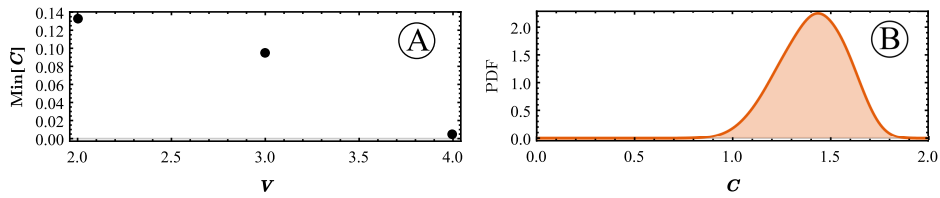}
\end{center}
\caption{(A) minimum value of $C$ found after numerical minimization as a function of the number of vertices $V$ taken into account in the foam. (B) Probability density function of the value of $C$ in the case $V=4$. The minimum value is $\min[C]=0.0056$ and the maximum value is $\max[C]=1.9998$.}
\label{fig:minimization_C_compil}
\end{figure}
Figure~\ref{fig:minimization_C_compil} show the minimum value of $C$ after numerical minimization of the cost function, and the probability density function to find a certain value of $C$. The second graph is computed with$10^5$ random spin-networks of the simplified model. A step of $0.1$ is used to compute the histogram.  We observe that very low values of $C$ are reachable, but these low values are not representative of general values of $C$ (the average value is $\approx 1.5$). This emphasizes that a specific design of $\psi_{bath}$ is required for the approximation. Due to many linear dependence between parameters, and the large number of parameters, the numerical optimization is not a difficult task, the convergence is fast. We also notice that the approximation is better for initial spinfoams of large dimension. This is explained by the fact that, the boundary of two distant vertices are weakly correlated.

\bibliographystyle{spphys}
\bibliography{biblio2}

\begin{thebibliography}{10}
\providecommand{\url}[1]{{#1}}
\providecommand{\urlprefix}{URL }
\expandafter\ifx\csname urlstyle\endcsname\relax
  \providecommand{\doi}[1]{DOI \discretionary{}{}{}#1}\else
  \providecommand{\doi}{DOI \discretionary{}{}{}\begingroup
  \urlstyle{rm}\Url}\fi

\bibitem{gambini_loops_1996}
R.~Gambini, J.~Pullin, A.~Ashtekar, \emph{Loops, {Knots}, {Gauge} {Theories}
  and {Quantum} {Gravity}}.
\newblock Cambridge {Monographs} on {Mathematical} {Physics} (Cambridge
  University Press, 1996).
\newblock \doi{10.1017/CBO9780511524431}

\bibitem{rovelli_quantum_2007}
C.~Rovelli, \emph{Quantum {Gravity}} (Cambridge University Press, Cambridge ;
  New York, 2007)

\bibitem{ashtekar_general_2014}
A.~Ashtekar, M.~Reuter, C.~Rovelli, arXiv:1408.4336  (2014).
\newblock \urlprefix\url{http://arxiv.org/abs/1408.4336}

\bibitem{rovelli_space_2018}
C.~Rovelli, arXiv:1802.02382  (2018).
\newblock \urlprefix\url{http://arxiv.org/abs/1802.02382}

\bibitem{rovelli_covariant_2014}
C.~Rovelli, F.~Vidotto, \emph{Covariant {Loop} {Quantum} {Gravity}: {An}
  {Elementary} {Introduction} to {Quantum} {Gravity} and {Spinfoam} {Theory}}
  (Cambridge University Press, 2014).
\newblock \doi{10.1017/CBO9781107706910}

\bibitem{rovelli_simple_2011}
C.~Rovelli, Journal of Physics: Conference Series \textbf{314}, 012006 (2011).
\newblock \doi{10.1088/1742-6596/314/1/012006}.
\newblock \urlprefix\url{http://arxiv.org/abs/1010.1939}.
\newblock ArXiv: 1010.1939

\bibitem{perez_spin-foam_2013}
A.~Perez, Living Rev. Relativ. \textbf{16}(1), 3 (2013).
\newblock \doi{10.12942/lrr-2013-3}.
\newblock \urlprefix\url{https://doi.org/10.12942/lrr-2013-3}

\bibitem{barrau_loop_2014}
A.~Barrau, J.~Grain, arXiv:1410.1714 [astro-ph, physics:gr-qc]  (2014).
\newblock \urlprefix\url{http://arxiv.org/abs/1410.1714}

\bibitem{cai_testing_2014}
Y.F. Cai, Y.~Wang, Physics Letters B \textbf{735} (2014).
\newblock \doi{10.1016/j.physletb.2014.06.019}.
\newblock \urlprefix\url{http://arxiv.org/abs/1404.6672}

\bibitem{perez_black_2017}
A.~Perez, Rep. Prog. Phys. \textbf{80}(12), 126901 (2017).
\newblock \doi{10.1088/1361-6633/aa7e14}.
\newblock \urlprefix\url{http://arxiv.org/abs/1703.09149}

\bibitem{christodoulou_transition_2017}
M.~Christodoulou, Transition de géométrie en gravité quantique à boucles
  covariante.
\newblock {PhD} {Thesis}, AIX-MARSEILLE Univeristy (2017).
\newblock \urlprefix\url{http://www.theses.fr/2017AIXM0273/document}

\bibitem{dambrosio_end_2020}
F.~D'Ambrosio, M.~Christodoulou, P.~Martin-Dussaud, C.~Rovelli, F.~Soltani,
  arXiv:2009.05016 [gr-qc]  (2020).
\newblock \urlprefix\url{http://arxiv.org/abs/2009.05016}.
\newblock ArXiv: 2009.05016

\bibitem{dona_numerical_2018}
P.~Dona, G.~Sarno, Gen Relativ Gravit \textbf{50}(10), 127 (2018).
\newblock \urlprefix\url{http://arxiv.org/abs/1807.03066}

\bibitem{dona_numerical_2019}
P.~Dona, M.~Fanizza, G.~Sarno, S.~Speziale, arXiv:1903.12624  (2019).
\newblock \urlprefix\url{http://arxiv.org/abs/1903.12624}

\bibitem{breuer_theory_2007}
H.P. Breuer, F.~Petruccione, \emph{The {Theory} of {Open} {Quantum} {Systems}}
  (Oxford University Press, Oxford, 2007)

\bibitem{feller_entanglement_2017}
A.~Feller, Entanglement and {Decoherence} in {Loop} {Quantum} {Gravity}.
\newblock {PhD} {Thesis}, ENS de Lyon (2017).
\newblock
  \urlprefix\url{https://tel.archives-ouvertes.fr/tel-01650029/document}

\bibitem{azouit_adiabatic_2016}
R.~Azouit, A.~Sarlette, P.~Rouchon, arXiv:1603.04630 [quant-ph]  (2016).
\newblock \urlprefix\url{http://arxiv.org/abs/1603.04630}

\bibitem{paetz_analysis_2010}
T.T. Paetz, An {Analysis} of the ‘{Thermal}-{Time} {Concept}’ of {Connes}
  and {Rovelli}.
\newblock {PhD} {Thesis}, Georg-August-Universität Göttingen (2010).
\newblock
  \urlprefix\url{http://www.theorie.physik.uni-goettingen.de/forschung/qft/theses/dipl/Paetz.pdf}

\bibitem{baez_introduction_1999}
J.C. Baez, arXiv:gr-qc/9905087  (1999).
\newblock \urlprefix\url{http://arxiv.org/abs/gr-qc/9905087}

\bibitem{livine_short_2011}
E.R. Livine, arXiv:1101.5061  (2011).
\newblock \urlprefix\url{http://arxiv.org/abs/1101.5061}

\bibitem{engle_lqg_2008}
J.~Engle, E.~Livine, R.~Pereira, C.~Rovelli, Nuclear Physics B
  \textbf{799}(1-2), 136 (2008).
\newblock \doi{10.1016/j.nuclphysb.2008.02.018}.
\newblock \urlprefix\url{http://arxiv.org/abs/0711.0146}

\bibitem{christodoulou_planck_2016}
M.~Christodoulou, C.~Rovelli, S.~Speziale, I.~Vilensky, Phys. Rev. D
  \textbf{94}(8), 084035 (2016).
\newblock \doi{10.1103/PhysRevD.94.084035}.
\newblock \urlprefix\url{https://link.aps.org/doi/10.1103/PhysRevD.94.084035}

\bibitem{lindblad_generators_1976}
G.~Lindblad, Commun.Math. Phys. \textbf{48}(2), 119 (1976).
\newblock \doi{10.1007/BF01608499}.
\newblock \urlprefix\url{https://link.springer.com/article/10.1007/BF01608499}

\bibitem{oeckl_renormalization_2006}
R.~Oeckl, The Tenth Marcel Grossmann Meeting pp. 2296--2300 (2006).
\newblock \doi{10.1142/9789812704030_0321}.
\newblock \urlprefix\url{http://arxiv.org/abs/gr-qc/0401087}

\bibitem{banburski_pachner_2015}
A.~Banburski, L.Q. Chen, L.~Freidel, J.~Hnybida, Phys. Rev. D \textbf{92}(12),
  124014 (2015).
\newblock \doi{10.1103/PhysRevD.92.124014}.
\newblock \urlprefix\url{http://arxiv.org/abs/1412.8247}

\bibitem{chen_bulk_2016}
L.Q. Chen, Phys. Rev. D \textbf{94}(10), 104025 (2016).
\newblock \doi{10.1103/PhysRevD.94.104025}.
\newblock \urlprefix\url{http://arxiv.org/abs/1602.01825}

\bibitem{dittrich_coarse_2016}
B.~Dittrich, E.~Schnetter, C.J. Seth, S.~Steinhaus, Phys. Rev. D
  \textbf{94}(12), 124050 (2016).
\newblock \doi{10.1103/PhysRevD.94.124050}.
\newblock \urlprefix\url{http://arxiv.org/abs/1609.02429}

\bibitem{feller_surface_2017}
A.~Feller, E.R. Livine, Class. Quantum Grav. \textbf{34}(4), 045004 (2017).
\newblock \doi{10.1088/1361-6382/aa525c}.
\newblock \urlprefix\url{http://arxiv.org/abs/1607.00182}

\bibitem{delcamp_towards_2017}
C.~Delcamp, B.~Dittrich, Class. Quantum Grav. \textbf{34}(22), 225006 (2017).
\newblock \doi{10.1088/1361-6382/aa8f24}.
\newblock \urlprefix\url{http://arxiv.org/abs/1612.04506}

\bibitem{gentle_regge_2002}
A.P. Gentle, General Relativity and Gravitation \textbf{34}(10), 1701 (2002).
\newblock \doi{10.1023/A:1020128425143}.
\newblock \urlprefix\url{https://doi.org/10.1023/A:1020128425143}

\bibitem{christodoulou_characteristic_2018}
M.~Christodoulou, F.~D'Ambrosio, arXiv:1801.03027 [gr-qc]  (2018).
\newblock \urlprefix\url{http://arxiv.org/abs/1801.03027}

\bibitem{garraway_nonperturbative_1997}
B.M. Garraway, Phys. Rev. A \textbf{55}(3), 2290 (1997).
\newblock \doi{10.1103/PhysRevA.55.2290}.
\newblock \urlprefix\url{https://link.aps.org/doi/10.1103/PhysRevA.55.2290}

\bibitem{barrett_lorentzian_2010}
J.W. Barrett, R.J. Dowdall, W.J. Fairbairn, F.~Hellmann, R.~Pereira, Class.
  Quantum Grav. \textbf{27}(16), 165009 (2010).
\newblock \doi{10.1088/0264-9381/27/16/165009}.
\newblock \urlprefix\url{http://arxiv.org/abs/0907.2440}

\bibitem{magliaro_regge_2013}
E.~Magliaro, C.~Perini, Int. J. Mod. Phys. D \textbf{22}(02), 1350001 (2013).
\newblock \doi{10.1142/S0218271813500016}.
\newblock \urlprefix\url{http://arxiv.org/abs/1105.0216}

\bibitem{han_asymptotics_2013}
M.~Han, M.~Zhang, Class. Quantum Grav. \textbf{30}(16), 165012 (2013).
\newblock \doi{10.1088/0264-9381/30/16/165012}.
\newblock \urlprefix\url{http://arxiv.org/abs/1109.0499}

\bibitem{livine_new_2007}
E.R. Livine, S.~Speziale, Phys. Rev. D \textbf{76}(8), 084028 (2007).
\newblock \doi{10.1103/PhysRevD.76.084028}.
\newblock \urlprefix\url{https://link.aps.org/doi/10.1103/PhysRevD.76.084028}

\bibitem{bianchi_coherent_2010}
E.~Bianchi, E.~Magliaro, C.~Perini, Phys. Rev. D \textbf{82}(2), 024012 (2010).
\newblock \doi{10.1103/PhysRevD.82.024012}.
\newblock \urlprefix\url{http://arxiv.org/abs/0912.4054}

\bibitem{rovelli_small_2018}
C.~Rovelli, F.~Vidotto, Universe \textbf{4}(11), 127 (2018).
\newblock \doi{10.3390/universe4110127}.
\newblock \urlprefix\url{http://arxiv.org/abs/1805.03872}

\bibitem{gross_superradiance:_1982}
M.~Gross, S.~Haroche, Physics Reports \textbf{93}(5), 301  (1982).
\newblock \doi{https://doi.org/10.1016/0370-1573(82)90102-8}.
\newblock
  \urlprefix\url{http://www.sciencedirect.com/science/article/pii/0370157382901028}

\bibitem{santos_master_2014}
J.P. Santos, F.L. Semião, Phys. Rev. A \textbf{89}(2), 022128 (2014).
\newblock \doi{10.1103/PhysRevA.89.022128}.
\newblock \urlprefix\url{http://arxiv.org/abs/1311.0018}

\bibitem{bianchi_entropy_2012}
E.~Bianchi, arXiv:1204.5122 [gr-qc, physics:hep-th]  (2012).
\newblock \urlprefix\url{http://arxiv.org/abs/1204.5122}

\bibitem{haroche_exploring_2006}
S.~Haroche, J.M. Raimond, \emph{Exploring the {Quantum}: {Atoms}, {Cavities},
  and {Photons}} (OUP Oxford, 2006)

\bibitem{louisell_quantum_1973}
W.H. Louisell, \emph{Quantum statistical properties of radiation} (John Wiley
  \& Sons Canada, Limited, 1973)

\bibitem{connes_von_1994}
A.~Connes, C.~Rovelli, Class. Quantum Grav. \textbf{11}(12), 2899 (1994).
\newblock \doi{10.1088/0264-9381/11/12/007}.
\newblock \urlprefix\url{https://doi.org/10.1088%2F0264-9381%2F11%2F12%2F007}

\bibitem{rovelli_statistical_1993}
C.~Rovelli, Class. Quantum Grav. \textbf{10}(8), 1549 (1993).
\newblock \doi{10.1088/0264-9381/10/8/015}.
\newblock \urlprefix\url{https://doi.org/10.1088%2F0264-9381%2F10%2F8%2F015}

\bibitem{menicucci_clocks_2011}
N.C. Menicucci, S.J. Olson, G.J. Milburn, arXiv:1108.0883 [gr-qc]  (2011).
\newblock \urlprefix\url{http://arxiv.org/abs/1108.0883}

\bibitem{gemmer_quantum_2009}
J.~Gemmer, M.~Michel, G.~Mahler, \emph{Quantum {Thermodynamics}: {Emergence} of
  {Thermodynamic} {Behavior} {Within} {Composite} {Quantum} {Systems}}, 2nd
  edn.
\newblock Lecture {Notes} in {Physics} (Springer-Verlag, Berlin Heidelberg,
  2009).
\newblock \urlprefix\url{https://www.springer.com/gp/book/9783540705093}

\bibitem{barrett_ponzano-regge_2009}
J.W. Barrett, I.~Naish-Guzman, Class. Quantum Grav. \textbf{26}(15), 155014
  (2009).
\newblock \doi{10.1088/0264-9381/26/15/155014}.
\newblock \urlprefix\url{http://arxiv.org/abs/0803.3319}

\end{thebibliography}

\end{document}